\documentclass[a4paper,twocolumn,english,preprint, aps]{revtex4-1}
\usepackage[T1]{fontenc}
\usepackage[latin9]{inputenc}
\usepackage{amsmath}
\usepackage{amssymb}

\makeatletter

\pdfpageheight\paperheight
\pdfpagewidth\paperwidth

\renewcommand*{\p@subsection}{}

\renewcommand*{\p@subsubsection}{}

\makeatother

\usepackage{babel}
\begin{document}

\preprint{Preprint to be Submitted}

\title{Dual Conformable Derivative\\
Definition, Simple Properties and Perspectives for Applications}

\author{Wanderson Rosa}
\email{wandersonfisica@outlook.com }

\affiliation{Universidade Federal Rural do Rio de Janeiro, UFRRJ-IM/DTL }

\address{Av. Governador Roberto Silveira s/n- Nova Iguaçú, Rio de Janeiro,
Brasil, 695014.}

\author{José Weberszpil}
\email{josewebe@gmail.com; josewebe@ufrrj.br}

\affiliation{Universidade Federal Rural do Rio de Janeiro, UFRRJ-IM/DTL }

\address{Av. Governador Roberto Silveira s/n- Nova Iguaçú, Rio de Janeiro,
Brasil - 695014.}

\date{\today}

\maketitle
In this communication, one shows that there exists in the literature
a certain form of deformed derivative that can here be identified
as the dual of conformable derivative. The deformed subtraction is
used here, together with the duality concept, as the basic definitions
and starting points in order to obtain the connected dual operators.
The q-exponential, in the context of generalized statistical mechanics,
is the eigenfunction of this dual conformable derivative. The basic
properties of the dual deformed-derivatives and also some perspective
of applications and simple models are presented. The importance of
this deformed derivative for position-dependent models is highlighted.
An outlook of potential applications and developments is presented.


\section{Introduction}

Over the last decades, several researchers have been pursuing alternative
formalisms and mathematical tools in order to describe complex systems
in a more reliable way. Complex systems are known to be composed by
a large number of simple members that mutually interact and can also
interact with their environment; for that reason they are considered
as open systems. They have the potential to give rise to macroscopic
new collective behavior and properties, including the manifestation
of new structures due to self-organization and internal-structure
interactions, new dynamics and possible macroscopic emergent properties.
Complex systems most likely involve nonlinear interactions between
sub- units with enhanced behavior of coherence or order that extends
far away from what could be reached to any individual sub-unit. That
is, there are long-range interactions.

The use of deformed derivative is justified here based on our proposition
that there exists an intimate relationship between dissipation, coarse-grained
media and some limit scale of energy for the interactions, as already
explained in Refs. \citep{weberszpil2015connection,weberszpil2016variational,weberszpil2017generalized,weberszpil2017structural,Jap-Weber-Sotolongo}.
In this context, deformed derivatives have emerged as a proposal to
deal with complex problems with the mathematical tools of local operators.

One of the most recent definitions of deformed derivative appeared
in 2014, with Khalil \citep{KHALIL201465}; it was called conformable
derivative. Another important kind of deformed derivative is the q-derived,
in the context of the nonextensive statistical mechanics \citep{borges2004possible}
and also fractal derivative \citep{CHEN20101754}, also called Hausdorff
derivative \citep{chen2006time,chen2017non}.

Here, one consider that physical basis involved in the justification
for the use of deformed derivatives, the mapping into the fractal
continuum \citep{weberszpil2015connection,balankin2012hydrodynamics,balankin2012map,balankin2016towards}.
We are not talking about including classical definitions of fractional
derivatives nor including operators of integer order acting on a d-dimensional
space, but one considers a mapping from a fractal coarse-grained (fractal
porous) space, which is essentially discontinuous in the embedding
Euclidean space, to a continuous one \citep{weberszpil2016variational}.
A mapping into a continuous fractal space naturally yields the need
for modifications in the derivatives and, with connection with the
metric, the modifications of the derivatives leads to a change in
the algebra involved, which, in turn may, lead arrive at a generalized
statistical mechanics with some suitable definition of entropy \citep{weberszpil2017generalized}.

Another explanation for deformed derivatives can be thought in terms
of a canonical transformation from one Euclidean space to a deformed
space \citep{da2018position}. A number of discussions on the physical
interpretation of deformed derivatives in terms of the Gateaux extended
derivative can be found in Ref. \citep{zhao2017general}.

To develop the dual conformable derivative, some concepts are necessary.
One of these is the concept of duality for derivative operators. The
basic ideas of duality, for q-deformed derivative operators, are originally
found in Ref. \citep{Phd,Phd-Borges}. But there, in terms of a property
instead of a basic definition. Also in the reference \citep{Phd},
appears a derivative of form $\tilde{D}_{x}^{\alpha}f=f^{\alpha-1}\dfrac{df}{dx}$,
where there was indicated that it has been first appeared, in this
form, in Ref. \citep{nobre2011nonlinear}. The latter reference does
not make clear the real origin of this derivative and was inserted
it in a certain \textit{ad hoc} mode. Also in Ref. \citep{Phd}, the
author shows that the dual derivative, related to its dual $\tilde{D}_{x}^{\alpha}f$,
is $D_{x}^{\alpha}f=x^{1-\alpha}\dfrac{df}{dx},$ indicating the origin
of the latter operator in the article of one of the authors here,
Weberszpil \citep{weberszpil2015connection}, but not making clear
the origin of both derivatives, $D_{x}^{\alpha}f$ and its dual $\tilde{D}_{x}^{\alpha}f$
and also, unfortunately, erroneously associated with fractional calculus,
that is a non-local formalism. The deformed chain rule was also indicated
there. The deformed algebra in those references is also different
from the one presented here.

In this contribution, inspired on the concepts of deformed operations,
duality and the definition of the conformable derivative (CD), one
shows that the deformed operator $\tilde{D}_{x}^{\alpha}f$ is in
fact a dual conformable derivative (DCD) operator. Also, by the deformed
chain rule, one shows that eigenvalue problems with q-exponential,
in the sense of generalizes statistical mechanics, can be more suitably
treated with the DCD operator.

Our Letter is outlined as follows. Section 2 addresses some mathematical
aspects to define the deformed subtraction and the $\tilde{D}_{x}^{\alpha}f$
operator. In Section 3, we focus on the duality concept and identification
of $\tilde{D}_{x}^{\alpha}f$ as the DCD operator and the dual conformable
integral. In Section 4, one obtains the deformed Leibniz product rule
and the deformed chain rule. Section 5 is devoted to some eigenfunction
and eigenvalue aspects and potential applications. Finally, in Section
6, we cast our general conclusions and possible paths for further
investigations.

\section{Conformable Subtraction}

In this section, motivated by the deformation on the argument of CD
\citep{KHALIL201465}, one obtains a deformed subtraction operation,
that one call Conformable Subtraction.

The CD is defined as \citep{KHALIL201465} 
\begin{equation}
D_{x}^{\alpha}f(x)=\lim_{\epsilon\rightarrow0}\frac{f(x+\epsilon x^{1-\alpha})-f(x)}{\epsilon}.\label{eq:Deformed Derivative-Deff}
\end{equation}

Note that the deformation is placed in the independent variable.

For differentiable functions, the CD can be written as

\begin{equation}
D_{x}^{\alpha}f=x^{1-\alpha}\dfrac{df}{dx}.\label{eq:conformable-deriv-differentiable}
\end{equation}
This can be obtained by a simple change of variable \citep{KHALIL201465}.

One now proceeds to briefly investigate two algebraic operations.

The argument of eq. (\ref{eq:Deformed Derivative-Deff}) is 
\begin{equation}
y=x+\epsilon x^{1-\alpha}.
\end{equation}
So, the infinitesimal parameter $\epsilon$ is 
\begin{equation}
\epsilon=\dfrac{y-x}{x^{1-\alpha}}.\label{eq:MotivAlgebra}
\end{equation}

Motivated by the generalized statistical mechanics approach and its
related algebra \citep{borges2004possible,nivanen2003generalized},
with the help eq.(\ref{eq:MotivAlgebra}), one can define a conformable
subtraction as 
\begin{equation}
y\ominus_{\alpha}x\equiv\dfrac{y-x}{x^{1-\alpha}}.
\end{equation}

With those simple algebraic definitions, the conformable derivative
and its dual conformable derivative, respectively, can be expressed
formally as

\begin{equation}
D_{x}^{\alpha}F\equiv\lim_{y\longrightarrow x}\dfrac{F(y)-F(x)}{y\ominus_{\alpha}x},
\end{equation}

\begin{equation}
\tilde{D}_{x}^{\alpha}F\equiv\lim_{y\longrightarrow x}\dfrac{F(y)\ominus_{\alpha}F(x)}{y-x}.
\end{equation}
For the DCD, it can be explicitly written as

\begin{alignat}{1}
\tilde{D}_{x}^{\alpha}F & =\lim_{y\longrightarrow x}\dfrac{F(y)-F(x)}{y-x}{[F(x)]}^{\alpha-1}=\nonumber \\
 & =F^{\alpha-1}\dfrac{dF}{dx}.\label{eq:Dual Conformable-First}
\end{alignat}

This latter expression first appeared in Ref. \citep{nobre2011nonlinear},
but without a clear origin.

\subsection*{Connections with Chen's fractal derivative}

It is worthy a short note here on a connection with Chen's fractal
derivative \citep{chen2006time,CHEN20101754,chen2017non}.

The Chen's fractal derivative is defined as

\begin{equation}
\frac{\partial g(x)}{\partial x^{\alpha}}=\lim_{x'\rightarrow x}\dfrac{g(x')-g(x)}{x\text{'}^{\alpha}-x^{\alpha}}.
\end{equation}
Now, by a simple variable change, 
\[
x'=x+\epsilon x^{1-\alpha}
\]
and consequently $\epsilon=\dfrac{x'-x}{x^{1-\alpha}}=x'\ominus_{\alpha}x.$

For $\epsilon\ll1$, we can writ, at first order in $\epsilon:$ 
\begin{equation}
x'^{\alpha}=x^{\alpha}\left[1+\epsilon x^{-\alpha}\right]^{\alpha}\approx x^{\alpha}\left[1+\alpha\epsilon x^{-\alpha}\right],
\end{equation}

\begin{equation}
x'^{\alpha}-x^{\alpha}\approx\alpha\epsilon=\alpha\left[\dfrac{x'-x}{x^{1-\alpha}}\right]=\alpha(x'\ominus_{\alpha}x).
\end{equation}
So,

\begin{alignat}{1}
\frac{\partial g(x)}{\partial x^{\alpha}} & =\lim_{x'\rightarrow x}\dfrac{g(x')-g(x)}{x\text{'}^{\alpha}-x^{\alpha}}\approx\nonumber \\
\lim_{x'\rightarrow x}\dfrac{g(x')-g(x)}{\alpha(x'\ominus_{\alpha}x)} & =\lim_{x'\rightarrow x}\dfrac{x^{1-\alpha}}{\alpha}\dfrac{F(x')-F(x)}{x'-x}=\nonumber \\
= & \dfrac{x^{1-\alpha}}{\alpha}\dfrac{dF}{dx}.
\end{alignat}

The result above shows that the Chen's fractal derivative is proportional
to CD for differentiable functions, up to the first order in the deformation
parameter $\epsilon.$

\section{Duality for Conformable Derivative and Conformable Integral}

The concept of duality for derivative operators is already present
in the standard calculus and it can be understood in a simple manner.

Consider the first order derivative of a real bijective continuous
differentiable function $y;$ $y,x\in\mathbb{R}.$

\begin{equation}
D_{x}^{1}y=\dfrac{dy}{dx},
\end{equation}

Now, considering an domain-image interchange as $x=x(y)$, we can
define the dual derivative of $x$ relative to $y$ as

\begin{equation}
\tilde{D_{y}^{1}}x=\dfrac{dx}{dy}.
\end{equation}

The following property will be considered as the main definition of
derivative duality: 
\begin{equation}
(\tilde{D}_{y}^{1}x(y))(D_{x}^{1}y(x))=1.\label{eq:Main-property-dual-integer}
\end{equation}
In this sense, this means that the first order order derivative and
its dual are self-dual \citep{Phd}. That is,

\begin{equation}
\dfrac{dy}{dx}\dfrac{dx}{dy}=1,
\end{equation}
or 
\begin{equation}
\dfrac{dx}{dy}=\left(\dfrac{dy}{dx}\right)^{-1}.
\end{equation}

Now, one can follow in the direction to extend this concept of duality
for the derivative operators and for a special case of deformed derivatives.
The conformable derivative is one kind of deformed derivatives of
a function $f$.

Considering the defining property (\ref{eq:Main-property-dual-integer})
extended to conformable derivative, the dual conformable-derivative
can be defined, considering one independent variable $x$ as function
of $y,$ where $x,y\in\mathbb{R},$ continuous and differentiable
functions:

\begin{equation}
\tilde{D_{y}^{\alpha}}x=\left[x^{1-\alpha}\dfrac{dy}{dx}\right]^{-1}=x^{\alpha-1}\dfrac{dx}{dy}.
\end{equation}
For a continuous differentiable real valued function $F,$ with $x$
as the independent variable, one can properly generalize the dual
conformable-derivative as 
\begin{equation}
\tilde{D_{x}^{\alpha}}F=F^{\alpha-1}\dfrac{dF}{dx},\label{eq:Dual Deriv differentiable}
\end{equation}
that corresponds to the expression of dual conformable-derivative
in eq. (\ref{eq:Dual Conformable-First}), for differentiable functions.

The expression in eq. (\ref{eq:Dual Deriv differentiable}) is identical
to eq. (\ref{eq:Dual Conformable-First}).

All the operators cast above are local operators and not related to
any version of non-integer fractional calculus; the latter is a non-local
operator. This is emphasized here in order to avoid any possible misunderstanding.

In short, one can cast the expressions for conformable-derivative
and its dual, for differentiable functions:

\begin{equation}
\begin{cases}
D_{x}^{\alpha}F=x^{1-\alpha}\dfrac{dF}{dx}, & (Conformable)\\
\tilde{D}_{x}^{\alpha}F=F^{\alpha-1}\dfrac{dF}{dx}. & (Dual-conformable)
\end{cases}
\end{equation}

An important point to be noticed here is that the deformations affect
different functional spaces, depending on the problem under consideration.
For the conformable derivative, the deformations are put in the independent
variable, which can be a space coordinate, in the case of, e.g, mass
position dependent problems, or even time or space-time variables,
for temporal dependent parameter or relativistic problems. The physics
will be main guide.

Deformed derivatives and deformed dual derivatives, in the context
of generalized statistical mechanics are also present. There, the
q-deformed derivative has also a dual derivative and a q-exponential
related function. For details the reader can see the Ref. \citep{da2018position}
and references therein. But here, the advantage for the use of dual-conformable-derivative
will be evident, particularly for issues associated with eigenvalue
problems and mass dependent problems.

\subsection*{The Dual Conformable-Integral}

In order to define a dual-conformable-integral, one recalls that the
conformable integral is already known to have the Riemann integral
structure \citep{KHALIL201465,weberszpil2016variational}, in such
a way that $D_{x}^{\alpha}(I_{x}^{\alpha}F)=F,$ where $I_{x}^{\alpha}F$
can be written as

\begin{equation}
I_{x}^{\alpha}F=\int_{\alpha}F(x)d^{\alpha}x.
\end{equation}

Here, the differential element $d^{\alpha}x$ is $d^{\alpha}x=x^{\alpha-1}dx.$

Following this reasoning, one can write an expression for the dual
conformable integral, in such a manner that the fundamental theorem
of calculus is also valid:

\begin{equation}
\tilde{D}_{x}^{\alpha}(\tilde{I}_{x}^{\alpha}F)=F.
\end{equation}

Defining the dual conformable integral as 
\begin{equation}
\tilde{I}_{x}^{\alpha}F=\int F^{1-\alpha}(x)F(x)dx,
\end{equation}
can be easily verified that

\begin{alignat}{1}
\tilde{D}_{x}^{\alpha}(\tilde{I}_{x}^{\alpha}F) & =\nonumber \\
= & F^{\alpha-1}(x)\dfrac{d}{dx}\left(\int F^{1-\alpha}(x)F(x)dx\right)\nonumber \\
= & F(x).
\end{alignat}

In so doing, we are ensuring the validity of the fundamental theorem
of calculus.

\section{The Leibniz Product and the Chain rules for Dual Conformable Derivative}

For conformable derivative, several properties are well known and
it is clear that they share similarities with the standard calculus
\citep{KHALIL201465}. So, the Leibniz product property and the chain
rule are, respectively,

\begin{equation}
D_{x}^{\alpha}(FG)=(D_{x}^{\alpha}F)G+(D_{x}^{\alpha}G)F,\label{eq: Leibniz Conformable}
\end{equation}
\begin{alignat}{1}
D_{x}^{\alpha}[F(G(x))] & =x^{1-\alpha}\dfrac{d[F(G(x))]}{dx}=\nonumber \\
= & x^{1-\alpha}\dfrac{dF}{dG}.\dfrac{dG}{dx}=\dfrac{dF}{dG}.(D_{x}^{\alpha}G).\label{eq:Chain Conformable}
\end{alignat}

For the dual-conformable derivative, the rules are directly obtained
as follows. For Leibniz product rule one has 
\begin{alignat}{1}
\tilde{D}_{x}^{\alpha}(FG) & =F^{\alpha-1}G^{\alpha-1}\dfrac{d(FG)}{dx}=\nonumber \\
= & F^{\alpha-1}G^{\alpha-1}\left[G\dfrac{dF}{dx}+F\dfrac{dG}{dx}\right],
\end{alignat}
that is symbolized as

\begin{equation}
\tilde{D}_{x}^{\alpha}(FG)=G^{\alpha}(\tilde{D}_{x}^{\alpha}F)+F^{\alpha}(\tilde{D}_{x}^{\alpha}G).\label{eq:Dual Leibniz}
\end{equation}

The chain rule is determined as

\begin{alignat}{1}
\tilde{D}_{x}^{\alpha}[F(G(x))] & =F^{\alpha-1}\dfrac{dF(G(x))}{dx}=\nonumber \\
= & F^{\alpha-1}\dfrac{dF}{dG}\dfrac{dG}{dx}=\dfrac{dG}{dx}(\tilde{D}_{x}^{\alpha}F).
\end{alignat}

That is, 
\begin{equation}
\tilde{D}_{x}^{\alpha}[F(G(x))]=\dfrac{dG}{dx}(\tilde{D}_{x}^{\alpha}F).\label{eq:Chain-Dual}
\end{equation}

This important result is of great valuable for eigenvalue problems.
Particularly with generalized statistical mechanics and with the so
called q-exponential functions.

Also important to note is that DCD is a nonlinear operator and in
this sense, 
\begin{alignat}{1}
\tilde{D}_{x}^{\alpha}[F+G] & =\left(\frac{F+G}{f.g}\right)^{\alpha-1}\left[G^{\alpha-1}\tilde{D}_{x}^{\alpha}F+F^{\alpha-1}\tilde{D}_{x}^{\alpha}G\right]\nonumber \\
\neq & \tilde{D}_{x}^{\alpha}F+\tilde{D}_{x}^{\alpha}G.
\end{alignat}
.

\section{Dual Conformable Derivative Eigenfuntion and Eigenvalue Problems}

The stretched exponential is an eigenfunction of the $D_{x}^{\alpha}$
operator \citep{weberszpil2015connection}, since the simple solution
of the equation

\begin{equation}
x^{1-\alpha}\dfrac{dy}{dx}=y,\label{eq:Conformable-equation}
\end{equation}
along with the the initial condition $y(0)=1$, leads to the solution
as an stretched exponential of form

\begin{equation}
y=\exp\left[\dfrac{x^{\alpha}}{\alpha}\right].
\end{equation}

Analogously, one may think of an equation for the dual-conformable-derivative
as

\begin{equation}
[F(x)]^{\alpha-1}\dfrac{dF(x)}{dx}=F(x).\label{eq:dual-eigen equation}
\end{equation}
Solving, along with the condition $F(0)=1,$ lead to the following
important function

\begin{equation}
F(x)=\left[1+(\alpha-1)x\right]^{1/(\alpha-1)}.
\end{equation}

This function is nothing but the reparametrized q-exponential, ubiquitous
in the Tsallis version of generalized statistical mechanics \citep{tsallis1988possible}.
This can clearly seen by redefining the relevant parameter $\alpha$
in term of the entropic parameter $q,$ as $\alpha-1=1-q$, or $\alpha=2-q$.
By this reparametrization, the solution of eq. (\ref{eq:dual-eigen equation})
becomes

\begin{equation}
F(x)=\left[1+(1-q)x\right]^{1/(1-q)},
\end{equation}
that is exactly the q-exponential \citep{borges2004possible}, $exp_{q}(x),$
in the Tsallis generalized statistical mechanics.

A very important point to stress here is that, in the case of dual-conformable-derivative
operator, the chain rule formalized in eq. (\ref{eq:Chain-Dual}),
allows eigenvalue problems. The q-derivative, a version used in the
context of Tsallis formalism, does not allow this. It can directly
be proven, with the help of eqs. (\ref{eq:Dual Deriv differentiable},
\ref{eq:Chain-Dual}), that the dual conformable-derivative applied
to an q-exponential function, with a dilatation on the independent
variable, leads to an eigenvalue problem, as follows:

\begin{equation}
\tilde{D}_{x}^{2-q}(e_{2-q}(\lambda x))=\lambda e_{2-q}(\lambda x),\label{eq:Eigen Value 2-q}
\end{equation}
where $e_{q}(\lambda x)$ is the q-exponential and $\lambda$ is some
real eigenvalue and we used $\alpha=2-q$ .

It can be verified that this not occur with the use of q-derivative,
unless $\lambda=1$ or if one use another kind of q-derivative, called
scale-q-derivative \citep{weberszpil2016variational,weberszpil2017structural}.

Another point to observe in eq. (\ref{eq:Eigen Value 2-q}) is the
appearance of $\alpha\rightarrow2-q$ duality. In terms of generalized
statistical mechanics, some aspects of this duality were studied in
Refs. \citep{naudts2002deformed,naudts2004generalized,wada2005connections}.

One more point to highlight here is that, despite this is the first
time that dual conformable-derivative is more clearly connected with
conformable derivative operator, the appearance of analogous derivative
is reported in Ref. \citep{nobre2011nonlinear}, however, in this
article, it was introduced in some an \textit{ad hoc} manner.

\subsection*{Potential Applications}
\begin{itemize}
\item Position-dependent Problems: A simple example is a model for a mass-dependent
harmonic oscillator. Using the dual conformable derivative definition,
one can deform Newtonian mechanics. For example, deforming the time
derivative, as follows: $x^{\alpha-1}\dfrac{d(\tilde{D}_{t}^{(\alpha)}x)}{dt}=-(\omega_{0,\alpha}^{2})x.$
Here, $\omega_{0,\alpha}$ is dimensionally consistent and related
to spring elastic constant and the mass of the oscillating point mass.
The solution to this equation can be obtained by proposing a q-exponential
kind of solution and rewriting it in terms of deformed circular functions.
It is worthy commenting that the solution to the present model is
different from the result of Ref. \citep{da2018position}, since here
the circular functions are simultaneously compatible with the q-exponential
and the dual conformable derivative. By other side, in the Ref. \citep{da2018position}
the solutions are compatible with q-dual deformed derivative and,
in this way, the models of deformed oscillators are different. 
\item Nonlinear Equations, Relativity and Quantum mechanics: Several possibilities
may be studied, like nonlinear Fokker-Planck equations, heat transfer,
and even nonlinear and position-dependent mass in quantum mechanics.
Also, several different problems can be revisited, like classical
fields \citep{godinho2012extending}, Bohmian mechanics \citep{helayel2012aspects},
Zitterbewegung problem \citep{weberszpil2014zitterbewegung} and so
on. 
\end{itemize}

\section{Conclusions and Outlook For Further Investigations}

In this work, motivated by the generalized statistical mechanics approach,
one build up a conformable subtraction and also used the duality definition
for derivative operators, in order to identify the deformed derivative
$\tilde{D}_{x}^{\alpha}f=f^{\alpha-1}\dfrac{df}{dx}$ as the dual
of conformable derivative. A number of properties have been discussed,
like deformed Leibniz and chain rules. The q-exponential, in the context
of generalized statistical mechanics, is the eigenfunction of this
DCD. The relevance and potentiality of DCD for position-dependent
and nonlinear problems have been put in evidence here. Inter-relations
with Chen's fractal derivative have also been treated in our contribution.

As a prospect for future research, the variational approach, similar
to what was done in Ref. \citep{weberszpil2016variational} but for
$\alpha$ near $1$ (low nonlinearity limit), shall be re-assed and
published elsewhere.

\textbf{Acknowledgments: }Thanks to J. A. Helayël-Neto for for reading
and reviewing the manuscript.

\bibliography{Dual-Conformable-10-05-18}

\end{document}